\documentclass[prl,twocolumn,superscriptaddress]{revtex4-2}
\usepackage{epsfig}
\usepackage{graphicx}
\usepackage{palatino}
\usepackage[english]{babel}
\usepackage{hyphenat}
\usepackage{amsmath}
\usepackage{amssymb}
\usepackage{mathtools}
\usepackage{mathrsfs}
\usepackage{slashed}
\usepackage{epstopdf}
\usepackage{xcolor}
\usepackage{booktabs}
\definecolor{lcolor}{rgb}{0.,0.0,0.}
\definecolor{citcolor}{rgb}{0,0.,0.5}
\usepackage[breaklinks,colorlinks,urlcolor=blue,citecolor=blue,linkcolor=blue]{hyperref}
\usepackage{multirow}
\usepackage{ltablex}

\newcommand{\beq}{\begin{eqnarray}}
\newcommand{\eeq}{\end{eqnarray}}

\newcommand{\bem}{\begin{multline}}
\newcommand{\eem}{\end{multline}}
\newcommand{\beg}{\begin{gather}}
\newcommand{\eeg}{\end{gather}}

\newcommand{\nn}{\nonumber\\}

\newcommand{\ben}{\begin{eqnarray*}}
\newcommand{\een}{\end{eqnarray*}}

\def\cO{{\cal O}}

\newcommand{\eqn}[1]{Eq.~\eqref{#1}}

\newcommand{\secn}[1]{Section~1}
\newcommand{\appn}[1]{Appendix~1}

\long\def\comment#1{ }

\def\med{\text{med}}

\def\and{\quad\text{and}\quad}

\newcommand{\rmd}{{\rm d}}

\newcommand{\rme}{{\rm e}}

\newcommand{\qhat}{\hat{q}}

\def\0{{\boldsymbol 0}}

\newcommand{\del}{\partial}

\newcommand{\abar}{\bar{\alpha}}

\begin{document}

\title{Jet Quenching Meets Gluon Saturation}

\author{Paul Caucal} 
\email{caucal@subatech.in2p3.fr}
\affiliation{SUBATECH UMR 6457 (IMT Atlantique, Universit\'e de Nantes, IN2P3/CNRS), 4 rue Alfred Kastler, 44307 Nantes, France}
\author{Kevin Eisenberg}
\email{eisenb72@msu.edu}
\author{Yacine Mehtar-Tani} 
\email{mehtartani@bnl.gov}
\affiliation{Physics Department, Brookhaven National Laboratory, Upton, NY 11973, USA}

\begin{abstract}
We present a theoretical framework for jet fragmentation in heavy-ion collisions based on the resummation of large energy logarithms. Exploiting the hierarchy of scales characteristic of jet quenching, we show that the jet function obeys the Banfi-Marchesini-Smye evolution equation, with medium-induced energy loss and color decoherence encoded in the initial condition. This structure reveals a close correspondence with saturation physics. In particular, the coherence angle emerges as the analog of the saturation scale and exhibits the same asymptotic scaling behavior under nonlinear evolution. As a proof of principle, we compute the jet nuclear modification factor to quantify the interplay between vacuum radiation and medium-induced color decoherence. Our framework provides a unified perturbative description of vacuum-like parton showers, medium-induced radiation, and color-coherence effects, paving the way for precision studies of jet quenching at RHIC and the LHC.

\end{abstract}

\maketitle

Jets produced in heavy-ion collisions provide a unique probe of the quark--gluon plasma, carrying information about the microscopic dynamics and transport properties of the strongly interacting medium~\cite{Blaizot:2015lma,Cao:2020wlm}. The increasing precision of jet measurements at RHIC and the LHC~\cite{Connors:2017ptx,Apolinario:2022vzg,Apolinario:2024equ}  calls for a comparable level of theoretical accuracy. A central lesson emerging from recent measurements is that jet suppression depends not only on the total jet energy but also on its internal substructure~\cite{STAR:2020ejj,STAR:2021kjt,ALargeIonColliderExperiment:2021mqf,ATLAS:2022vii,CMS:2024zjn}. This behavior reflects the phenomenon of color coherence: the medium resolves the internal color structure of a jet only above a characteristic coherence angle $\theta_c$~\cite{Mehtar-Tani:2010ebp,Mehtar-Tani:2011hma,Mehtar-Tani:2011vlz,Casalderrey-Solana:2011ule,Mehtar-Tani:2011lic,Mehtar-Tani:2012mfa,Casalderrey-Solana:2012evi}. Understanding the dynamics governing this scale has therefore become one of the central challenges in jet-quenching theory ~\cite{Mehtar-Tani:2016aco,Hulcher:2017cpt,Caucal:2018dla,Casalderrey-Solana:2018wrw,Caucal:2019uvr,Caucal:2021cfb,Pablos:2022mrx,Cunqueiro:2023vxl,Casalderrey-Solana:2019ubu,Mehtar-Tani:2021fud,Takacs:2021bpv,Attems:2022otp}.

While the qualitative picture of medium-induced energy loss is now well established, systematically improvable perturbative calculations remain limited. Current phenomenological descriptions rely primarily on Monte Carlo event generators~\cite{Zapp:2008gi,Zapp:2012ak,Zapp:2013vla,Schenke:2009vr,Majumder:2013re,Wang:2013cia,Casalderrey-Solana:2014bpa,Caucal:2018ofz,Putschke:2019yrg,Karpenko:2024fgg,Zapp:2026cqf} or fixed-order calculations supplemented by partial logarithmic resummations. Unlike modern precision jet physics in the vacuum, however, no framework currently exists that systematically resums the soft-collinear radiation~\cite{Becher:2003qh,Bauer:2011uc,Larkoski:2015zka,Becher:2015hka,Becher:2016mmh} responsible for color coherence while preserving a factorized description of jet-medium interactions.

Recent developments based on effective field theory have established a factorization framework for inclusive jet production in heavy-ion collisions~\cite{Mehtar-Tani:2024smp,Mehtar-Tani:2025xxd,Vaidya:2026yfa}. Much like high-energy factorization at small $x$, this formalism separates short-distance perturbative dynamics from long-distance matrix elements encoding the interaction with the medium~\cite{Qiu:2019sfj}. The remaining missing ingredient is the all-order evolution of the collinear-soft sector, whose dynamics governs the transport of energy outside the jet cone and encodes the interplay between vacuum radiation and medium-induced color decoherence.

In this Letter we solve this problem by showing that the collinear-soft function obeys the Banfi--Marchesini--Smye (BMS) evolution equation~\cite{Banfi:2002hw,Dasgupta:2001sh}. Medium-induced radiation enters exclusively through the initial condition of this evolution, providing a factorized description in which the medium determines the boundary condition while vacuum evolution resums the large logarithms $\ln(E/\varepsilon)$ associated with energy $\varepsilon$ radiated outside the jet. This establishes, for jet quenching, a perturbative framework analogous to high-energy factorization in saturation physics, where nonlinear Balitsky--Kovchegov (BK) and Jalilian-Marian--Iancu--McLerran--Weigert--Leonidov--Kovner (JIMWLK) evolution ~\cite{Balitsky:1995ub,Kovchegov:1999yj,Jalilian-Marian:1997qno,Jalilian-Marian:1997jhx,Kovner:2000pt,Iancu:2000hn,Iancu:2001ad,Ferreiro:2001qy}  acts upon an initial condition provided by the McLerran--Venugopalan (MV) model~\cite{McLerran:1993ni,McLerran:1993ka}.

The correspondence extends far beyond a formal analogy. It reveals that the coherence angle in jet quenching and the saturation scale in small-$x$ QCD play precisely analogous roles: although they arise in different physical settings, both constitute emergent measures of color correlations in angular and transverse-coordinate space, respectively. As a consequence, several remarkable phenomena previously associated with nonlinear small-$x$ evolution, including traveling-wave solutions, geometric scaling, and universal asymptotic evolution, naturally emerge in the jet-quenching problem.

Building on this correspondence, we derive the energy dependence of the coherence angle from BMS evolution and compute the nuclear modification factor, including for the first time the combined effects of vacuum collinear-soft evolution and medium-induced color decoherence. Our results establish a unified perturbative framework connecting jet quenching, non-global logarithms~\cite{Dasgupta:2001sh}, and gluon saturation, providing a systematic foundation for precision calculations of jet observables in heavy-ion collisions.

\section{QCD Evolution of Jet Energy Loss}

\begin{figure}
    \centering
    
    \includegraphics[width=0.7\linewidth]{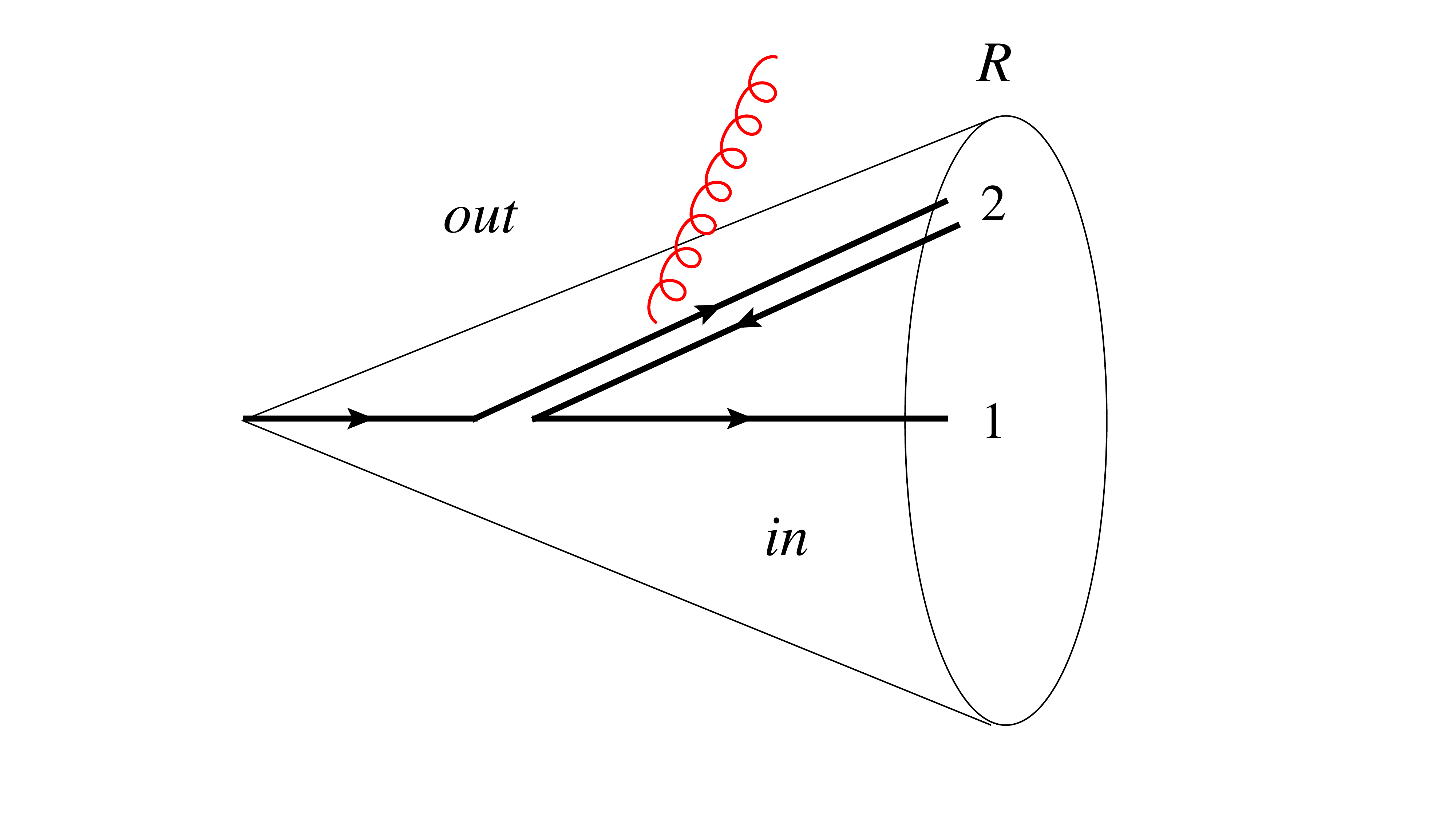}
    \caption{Illustration of the collinear-soft evolution in the large $N_c$ and $R\ll 1$ limits described by the BMS equation.}
    \label{fig:BMS-cartoon}
\end{figure}

The emergence of the BMS equation in the context of jet quenching will play a central role in establishing a connection between the physics of color (de)coherence and gluon saturation. This connection is reflected, in part, through a direct analogy between the coherence angle and the saturation scale. This correspondence becomes manifest in the threshold regime of inclusive jet production \cite{Dai:2017dpc,Liu:2017pbb,Vaidya:2026yfa}, where the dynamics of energy transport outside the jet cone of opening angle $R$ is described by a collinear-soft function that plays the same role as the dipole $S$-matrix in high-energy scattering. 

Our starting point is the factorization formula presented in \cite{Mehtar-Tani:2024smp,Mehtar-Tani:2025xxd}. At leading logarithmic accuracy and leading power in the threshold $1/N$ expansion, the inclusive jet cross section factorizes as
\begin{align}\label{eq:factorization-LL}
\frac{\rmd\sigma_{\rm jet}}{\rmd p_T}
=\sum_{i=q,g} H_{i}(p_T,\mu)
\, 
C_i(\mu) \, S_{i}(\mu,\nu=N/p_T)\,,
\end{align}

where $\mu$ is the factorization scale, eventually chosen $\mu\sim p_T$, and the collinear-soft function in Laplace space 
\begin{align}
S_{i}(\mu,\nu)
=
\int_0^{\infty}\rmd \varepsilon \ e^{-\nu\varepsilon}
S_{i}(\mu,\varepsilon)\,.\label{eq:cs-factor-laplace}
\end{align}

The hard function $H_{i}$ describes the short-distance production of a quark or gluon. It falls steeply with transverse momentum, $H(p_T)\sim p_T^{-N},$ with~$N\gg 1$. $C_i$ denotes the hard-collinear matching coefficient. 

$S_i(\mu,\nu)$, on the other hand, admits a representation in terms of lightlike Wilson lines and is the direct analog of the dipole $S$-matrix appearing in high-energy factorization \cite{Weigert:2003mm,Caron-Huot:2015bja}. $S_i(p_T,\varepsilon)$ admits the interpretation of a probability distribution for a parton of flavor $i$ and transverse momentum $p_T+\varepsilon$ to radiate an energy $\varepsilon$ outside the jet cone. 
Additional details on the derivation of \eqn{eq:factorization-LL} are provided in the Supplemental Material.

As demonstrated in~\cite{Banfi:2002hw,Weigert:2003mm,Caron-Huot:2015bja} and in a companion paper~\cite{Mehtar-Tani:2026}, the function $S_i$ satisfies the BMS equation~\cite{Banfi:2002hw}, which, in the large $N_c$ limit and for~$R\ll1$, reduces to the coupled set of equations:

\begin{widetext} 
\begin{align}\label{eq:bms-1}
\frac{\rmd S_{1}(\mu,\nu)}{\rmd \ln \mu}&=   \frac{\abar}{2}\left( -\ln \frac{4e^{2\gamma_E}\mu^2 \nu^2}{R^2}+\ln\left[1-\frac{\boldsymbol{\theta}_1^2}{R^2}\right]\right)  S_{1}(\mu,\nu) + \frac{\abar}{2\pi} \int_{\theta_{2}\le R} \frac{\rmd^2 \boldsymbol{\theta}_{2}}{\boldsymbol{\theta}_{12}^2}   \left[S_{2}(\mu,\nu)S_{12}(\mu,\nu)-S_{1}(\mu,\nu) \right]\,,\\
\frac{\rmd S_{12}(\mu,\nu)}{\rmd \ln \mu}&=  - \frac{\abar}{2}  \ln\left(1+\frac{R^2\boldsymbol{\theta}_{12}^2}{(\boldsymbol{\theta}_1^2-R^2)(\boldsymbol{\theta}_2^2-R^2)}\right)  S_{12}(\mu,\nu) + \frac{\abar}{2\pi} \int_{\theta_3\le R} \rmd^2 \boldsymbol{\theta}_3\frac{\boldsymbol{\theta}_{12}^2}{\boldsymbol{\theta}_{13}^2\boldsymbol{\theta}_{32}^2}   \left[S_{13}S_{32}-S_{12} \right] \,.\label{eq:bms-12}
\end{align}
\end{widetext} 
Here $S_1$, $S_{12}$ respectively describes the energy loss probability distribution (in Laplace space) by a single prong or a dipole. In the large $N_c$ limit, $S_1$ and $S_{12}$ are the only degrees of freedom, with $S_q=S_1$ and $S_g=S_1^2$. 
$S_1$ implicitly depends on the two-dimensional angular vector $\boldsymbol{\theta}_1$, with magnitude $\theta_1=|\boldsymbol{\theta}_1|$ and azimuthal orientation given by the emission direction. Likewise, $S_{12}$ depends on the angular vectors $\boldsymbol{\theta}_1$ and $\boldsymbol{\theta}_2$ associated with the two legs of the antenna. We adopt the shorthand notation $\boldsymbol{\theta}_{ij}\equiv\boldsymbol{\theta}_i-\boldsymbol{\theta}_j$.
The BMS equation can also 
be formulated in terms of the ``time'' evolution variable $\tau=\bar\alpha_s\ln(\mu\nu)$ with $\abar=\alpha_s N_c/\pi$ (for fixed strong coupling $\alpha_s$). The physical picture of this evolution, sketched in Fig.\,\ref{fig:BMS-cartoon} is that of a sequence of soft emissions inside the jet, which in the large $N_c$ limit reduces to a sequence of dipole splittings, followed by a final one 
(represented by a gluon in Fig.\,\ref{fig:BMS-cartoon}) outside the jets which contributes to its energy loss $\sim \varepsilon$. The evolution is strongly ordered in energy within the logarithmic range $p_T\gg \mu\gg p_T/N\sim \varepsilon$. The evolution of $S_1$ describes both the energy loss of the jet as a total color charge through the first term in parentheses, also computed in~\cite{Dai:2017dpc}, and non-global contributions from resolved soft-collinear intrajet splittings encoded in the mixing with $S_{12}$.

Since $\nu$ is eventually evaluated in Eq.\,\eqref{eq:factorization-LL} at the characteristic scale $N/p_T$, the BMS evolution effectively resums powers of $\alpha_s\ln N$ to all orders. This $\alpha_s\ln N$ resummation is analogous to the standard threshold resummation encountered in inclusive jet spectra~\cite{Dai:2017dpc,Liu:2017pbb}, and should be contrasted with the $\alpha_s\ln R$ resummation~\cite{Dasgupta:2014yra,Dasgupta:2016bnd,Kang:2016mcy,Dai:2016hzf} arising in the regime $R\ll 1$, which is governed by the DGLAP evolution of the collinear jet function. In the BMS approach, powers of $\varepsilon/p_T\sim 1/N$ are neglected, while the evolution keeps track of power corrections in $\theta/R$ where $\theta$ generically denotes the relative angle of emission, e.g.~$\theta_1,\theta_{12}$. The collinear $\ln(R)$ resummation, on the other hand, is all power in $\varepsilon/p_T$ but leading power in $\theta/R$ only. The BMS equation therefore provides a more precise treatment of the geometry of intrajet cascades. As we will discuss in the concluding section, this improvement becomes particularly important when accounting for color-coherence effects in the medium. For phenomenologically relevant values, with $N\sim 5\, \text{--}\, 6$ at the LHC and $N\sim 8 \,\text{--}\, 10$ at RHIC, one finds that $\ln N$ is numerically larger than $\ln(1/R)$ for $R\sim 0.4$.

In vacuum, the initial condition to the BMS equation is evidently $S_1(\tau=0)=1$ and $S_{12}(\tau=0)=1$, such that, after inverse Laplace transform, the probability distribution of losing an energy $\varepsilon$ without radiation is a $\delta(\varepsilon)$ function. Likewise, for the two-prongs energy loss, we have $S_{12}(\tau,\theta_{12}=0)=1$ at any $\tau$ as an antenna of vanishing opening angle cannot radiate.

\section{Color Decoherence as an Initial Condition}

Because the medium scale {\it lives} at the lower boundary of the logarithmic phase space $p_T \gg \omega \gg p_T/N$, at this accuracy the medium is encoded in the initial condition of the BMS evolution, in close analogy with the MV model which provides the initial condition for small-$x$ evolution in high-energy scattering. Independently of its microscopic realization, this initial condition should contain two characteristic scales: an energy scale associated with medium-induced energy loss and an angular scale, $\theta_c$, associated with the medium's ability to resolve the jet's internal structure. The latter underlies the phenomenon of color  decoherence and plays a role analogous to the saturation scale in small-$x$ physics.

While the correspondence between the BMS equation in vacuum and the BK equation has been extensively discussed in the literature~\cite{Hatta:2013iba,Weigert:2003mm}, it has remained largely formal. In the vacuum, the BMS equation governs the evolution of Wilson-line correlators evaluated in the vacuum state, whereas in saturation physics the BK equation describes the evolution of Wilson-line correlators averaged over the non-trivial color fields of a hadronic target. In this Letter, we show that this correspondence extends naturally to jet quenching, where the quark--gluon plasma provides the non-trivial environment that determines the initial condition for the BMS evolution. As a consequence, the emergent coherence angle plays a role directly analogous to the saturation scale, with both quantities arising as dynamical measures of color correlations generated by their respective environments.

To illustrate the emergence of these scales, it is instructive to evaluate $S_1$ and $S_{12}$ perturbatively in the limit of a large QGP, where a kinetic description applies and multiple medium-induced emissions and scatterings can be treated as independent processes. We adopt this analytically tractable framework to expose the physical content of the evolution equations. The resulting picture is, however, considerably more general and does not depend on the detailed modeling of the medium interactions. Rather, it relies only on the assumption that medium-induced energy loss is controlled by gluons with characteristic energies of order $p_T/N$, i.e.~by emissions that reside at the infrared boundary of the vacuum BMS evolution and therefore naturally determine its initial condition.

In this approximation, the plasma is described by the transverse momentum broadening coefficient $\hat q$.
To be more concrete, let us use the BDMPS-Z result~\cite{Baier:1996kr,Baier:1996sk,Zakharov:1996fv,Zakharov:1997uu} for the medium-induced gluon rate $\Gamma(\omega)=\bar\alpha_s\sqrt{\qhat/\omega^3}$ from a parton propagating over a distance $L$. For a single emission with energy $\omega$ emitted in the out-region (medium-induced gluons are typically emitted at large angles~\cite{Blaizot:2014ula}), $S_{1}^{(1)}(p_T=\nu^{-1},\nu) =-\Gamma(\nu) L $ with
\begin{align}
      \Gamma(\nu)= \int_0^{\infty}\rmd \omega \  \Gamma(\omega)\left(1- \rme^{-\omega \nu}\right)\,,\label{eq:S1med-LO}    
\end{align}
where the $1$ term 
comes from virtual emissions. 
The $\omega^{-3/2}$ shape of the rate is such that the integral over gluon energies is controlled by the lower boundary $\nu^{-1}\sim p_T/N$ and is therefore clearly not logarithmic in the phase space $p_T/N\ll \omega \ll p_T$. To leading logarithmic accuracy, medium-induced gluon energy loss and vacuum energy loss have no overlap in phase space, and therefore the two mechanisms simply factorize in Laplace space. This justifies the use of the medium-induced energy loss distribution as an initial condition to the BMS equation.

Exponentiating independent emissions  yields~\cite{Baier:2001yt}
\begin{equation}\label{eq:P1med}
        S_{1,\rm med}(\nu)=e^{-\Gamma(\nu) L}\,,
    \end{equation}
where the radiation rate in this approximation yields $ \Gamma(\nu)\simeq 2\sqrt{\pi\bar\alpha_s^2\qhat \nu }$. 

This expression assumes that all radiated gluons are emitted at large angles and is therefore independent of the emission angle $\boldsymbol{\theta}_1$. This approximation can be relaxed through a more realistic treatment of the geometry of the radiation spectrum. Here, to isolate the evolution dynamics, we neglect such geometric effects and assume that all radiated energy is lost outside the jet. A more refined treatment would introduce additional soft scales, potentially enlarging the phase space available for BMS evolution.

\begin{figure}
    \centering
    \includegraphics[width=0.95\linewidth,page=1]{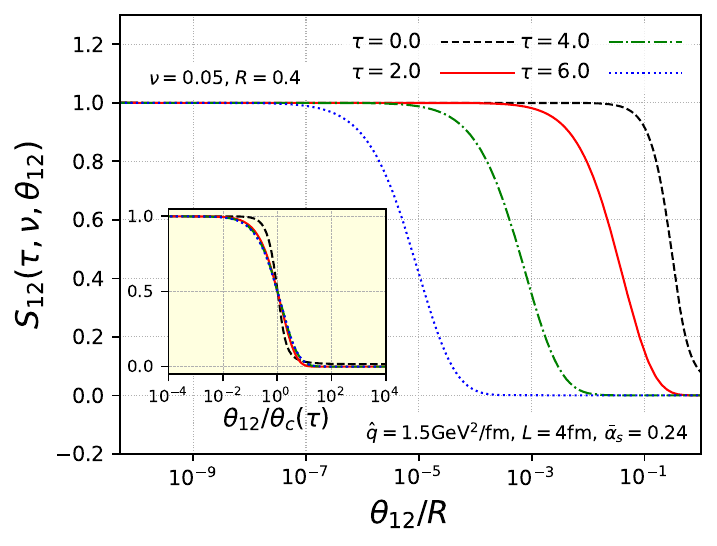}
    \caption{BMS evolution of the antenna energy-loss initial condition for several evolution times $\tau$. The solutions develop traveling-wave behavior and exhibit geometric scaling when plotted as a function of the rescaled angle $\theta_{12}/\theta_c(\tau)$, as shown in the inset.}
    \label{fig:2prong-BMS}
\end{figure}

The crucial ingredient is the dipole initial condition which takes the form \cite{Mehtar-Tani:2017ypq}
  
\begin{align}
\label{eq:main-result}
&S_{12,\rm med}(\nu) =   S_1(\nu, L) \, S_2(\nu, L)\\
 &- 2 \int_0^L \rmd t  \, S_1(\nu, L-t) \, S_2(\nu, L-t)  \Big[ 1- \Delta_\med(t) \Big]  \, \Gamma (\nu)\,.\nonumber
\end{align}

The medium decoherence parameter
$\Delta_{\rm med}(t)$
interpolates between coherent and incoherent energy loss.
For a dense medium,
\begin{align}
\Delta_{\rm med}(t)
=
1-\exp\!\left[
-\frac{1}{12}\hat q\,\theta_{12}^2 t^3
\right],
\end{align}
which introduces the characteristic angle $\theta_c
=2(\hat q L^3)^{-1/2}$. 

Dipoles with
$\theta_{12}\ll\theta_c$
remain color coherent and lose energy as a single charge. Using the rate equation $\del S_1/\del t = \Gamma\, S_1 $ we can show that $S_{12}(\nu)\to 1$ such that $S_{12}(\varepsilon)=\delta(\varepsilon)$ in physical space. In the opposite limit,
dipoles with
$\theta_{12}\gg\theta_c$ lose energy independently and $S_{12,\rm med}$ is then the product of two independent energy loss distribution $S_{1} S_2$. 

The $\theta_{12}$ dependence of $S_{12}$ is represented by the black dashed line in Fig.\,\ref{fig:2prong-BMS}. This behavior is akin to that of the dipole $S$-matrix $S(r_\perp)$ of a color dipole of fixed transverse size $r_\perp$ in the MV model:
for $r_\perp\ll 1/Q_s$ with $Q_s$ the nucleus saturation scale, we have $S(r_\perp)=1$ by color transparency, while for $r_\perp\gg 1/Q_s$, we have $S(r_\perp)\to 0$ in the strong scattering regime.

\section{Coherence Angle as the Saturation Scale}

It is actually fruitful to pursue this analogy further, as it allows us to translate into the present context several interesting emergent physical phenomena that have been discussed primarily in the framework of small-$x$ physics. In particular, it is well known that the BK evolution of $S(r_\perp)$ admits traveling waves solutions~\cite{Munier:2003vc,Munier:2003sj}, with the front velocity, corresponding to $\rmd \ln Q_s^2(x)/\rmd \ln(x)$ given asymptotically by a constant $\simeq 4.88\abar$~\cite{Mueller:2002zm}. The physical picture that has emerged from the previous sections is that the energy-loss distribution is determined, in Laplace space, by the BMS evolution of an initial condition set by the medium. The spacelike-timelike duality~\cite{Mueller:2018llt} between the BK and BMS equations~\cite{Marchesini:2003nh,Weigert:2003mm,Hatta:2008st,Avsar:2009yb}, realized through stereographic projection, provides the foundation for this analysis.

A first striking consequence of this duality is the existence of traveling wave solutions in the jet quenching problem. In order to highlight this phenomenon, we have solved Eq.\,\eqref{eq:bms-12} using Eq.\,\eqref{eq:main-result} as an initial condition. The $\theta_{12}$ dependence of the result, for a fixed $\nu=0.05$ is shown Fig.\,\ref{fig:2prong-BMS} for several values of $\tau$. As the effective time $\tau$ increases, the Laplace transform of the probability distribution to lose energy in the out-of-jet region behaves like a traveling wave whose front, initially set by the shape of the medium boundary condition, is propagating towards smaller angles. Like in the saturation problem, $S_{12}$ exhibits \textit{geometric scaling}~\cite{Stasto:2000er,Iancu:2002tr,Kwiecinski:2002ep}, meaning that at large $\tau$, it becomes a function of $\theta_{12}/\theta_c(\tau)$ where the critical angle $\theta_c(\tau)$ for $\tau>0$ is conventionally defined as the angle such that $S_{12}(\tau,\nu)=1/2$; this is illustrated in the inset plot. 

Another remarkable aspect of the coherence-saturation correspondence is that it determines the $\tau=\bar\alpha_s \ln(p_T\nu)$ dependence of $\theta_c$ induced by quantum corrections. In vacuum, the critical angle, which marks the lower edge of the "buffer" region~\cite{Dasgupta:2002bw} where real soft-collinear emissions are suppressed and thus sets the effective angular size of the jet, scales as $R e^{-c\tau}$ with $c\simeq 2.44$~\cite{Neill:2016stq}. The medium leads to a memory loss effect of the initial size $R$ of the jet, such that  
jet cone angle is replaced by the medium coherence angle from the initial condition:
\beq 
R\quad \to\quad \theta_c(0)\sim \frac{1}{\sqrt{\qhat L^3}}\,.
\eeq
The exact analytic asymptote based on the duality between the BK and BMS equations is
\begin{align}\label{eq:log-thetac}
    \ln(\theta_c(\tau))=-c\tau+b\ln(\tau)+\textrm{const.}+\mathcal{O}(\tau^{-1/2})\,,
\end{align}
with $c\simeq 2.44$ and $b\simeq 1.20$~\cite{Neill:2016stq}.  
This shows that the energy dependence of the coherence angle belongs to the same universality class as the saturation scale, with the medium entering only through the initial condition.
At asymptotically large energies, corresponding to $\tau=\bar\alpha_s\ln(N)\to \infty$, the dynamics  of the buffer region becomes universal and controlled by vacuum physics, but the overall normalization encoded in the constant term is set by the medium critical angle $\theta_c(0)\sim(\qhat L^3)^{-1/2}$ instead of $R$. This observation constitutes the central physics result of this work.

Table~\ref{tab:dictionary} summarizes the correspondence between decoherence and saturation physics, highlighting the common structure of the two nonlinear QCD problems underlying high-energy factorization and collinear-soft factorization in jet quenching.

\begin{widetext}
    
\begin{table}[t]
\centering
\small
\renewcommand{\arraystretch}{1.6}
\begin{tabular}{lcc}
\hline\hline
 & BMS evolution in a medium \quad &\quad BK evolution at small-$x$ \\ 
\hline 
Wilson-line correlator &
$S_{12}(\tau,\theta_{12})$
&
$S_{12}(Y,r_\perp)$
\\

Evolution variable &
$\tau=\bar\alpha_s\ln (p_T/\varepsilon)$
&
$Y=\bar\alpha_s\ln(1/x)$
\\

Initial condition &
Energy loss
&
MV model
\\

Emergent scale &
$\theta_c$
&
$Q_s^{-1}$
\\

Scaling regime &
$\theta_{12}/\theta_c(\tau)$
&
$r_\perp Q_s(Y)$
\\[3pt]

\hline\hline
\end{tabular}
\caption{Correspondence between medium-modified BMS evolution and nonlinear small-$x$ evolution.}
\label{tab:dictionary}
\end{table}

\end{widetext}

\section{Phenomenological Consequences}

\begin{figure}
    \centering
    \includegraphics[width=0.95\linewidth]{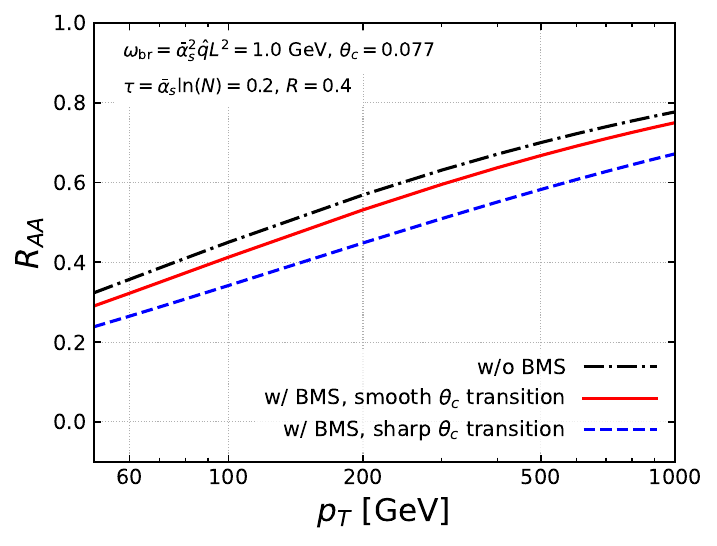}
    \caption{Nuclear modification factor for quark jets comparing the single-parton approximation with the parton shower described by the BMS equation, including color coherence effects. We observe substantial variations arising from different implementations of the color coherence angle.} 
    \label{fig:RAA}
\end{figure}

Beyond the conceptual correspondence between jet quenching and small-$x$ saturation, we now turn to the phenomenological consequences of BMS evolution for jet quenching at RHIC and the LHC. A central motivation for this analysis is that the theoretical description of the transition between coherent and decoherent medium-induced energy loss remains a major source of uncertainty in current phenomenological studies~\cite{Mehtar-Tani:2021fud}. First, our calculation predicts a 10--25\% relative variation in $\theta_c$ due to BMS evolution for $\tau\sim 0.2$--$0.5$ from LHC to RHIC, underscoring the phenomenological importance of its scale dependence. Further, existing calculations of inclusive jet suppression~\cite{Caucal:2019uvr,Mehtar-Tani:2021fud,Pablos:2025cli} resum the leading logarithms $\ln(R/\theta_c)$ but neglect power corrections in $\theta_c/R$, thereby describing the coherence transition only in the strict logarithmic approximation.  
We show that the BMS formalism provides the first systematic treatment of this transition and leads to sizeable phenomenological effects associated with its finite-width structure.

As a proof-of-concept study, we compute the nuclear modification factor, as obtained by dividing Eq.\,\eqref{eq:cs-factor-laplace} computed using the medium initial conditions with the same expression using the vacuum initial condition. We use $\tau=0.2$ corresponding to $\bar\alpha_s=0.12$ and $N=5$. The result of this calculation, shown in red in Fig.\,\ref{fig:RAA}, is first compared with a scenario in which the BMS vacuum evolution is switched off, meaning using directly $R_{AA}(p_T)=S_{1,\rm med}(N/p_T)$ (black dash-dotted line in Fig.\,\ref{fig:RAA}). One observes that the $\bar\alpha_s\ln(N)$ resummation leads to an additional suppression, which is attributed to the increase in the number of resolved intrajet partons arising from soft-collinear emissions in the inside-jet region. The other scenario, shown in blue in Fig.\,\ref{fig:RAA} consists in changing the initial condition for $S_{12,\rm med}$ by replacing Eq.\,\eqref{eq:main-result} with a sharp transition at the critical angle $\theta_c$, namely,
$ S_{12,\rm med}(\nu)=\Theta(\theta_{c}-\theta_{12})+S_{1,\rm med}^2(\nu)\Theta(\theta_{12}-\theta_c)$,
a simplified model for color decoherence  
used in Monte-Carlo implementation of in-medium parton shower~\cite{Caucal:2019uvr} or in semi-analytic results based on collinear DGLAP evolution~ 
\cite{Mehtar-Tani:2021fud,Mehtar-Tani:2017web,Mehtar-Tani:2024jtd,Pablos:2025cli}. This scenario allows us to investigate to what extent a more precise treatment of the $\theta_c$ transition, which is possible thanks to the formalism proposed in this Letter, is phenomenologically relevant. Fig.\,\ref{fig:RAA} shows that, for typical values of $\tau\sim 0.2$ at the LHC, the effect of power corrections in $\theta_c/R$ encoded in the smooth $S_{12,\rm med}$ are $\mathcal{O}(20\%)$, while the BMS evolution yields finally an $\mathcal{O}(10\%)$ effect on $R_{AA}$. At RHIC, where $\tau \sim 0.4$, these effects are expected to be more pronounced owing to the lower jet transverse momentum ($p_T \sim 20$ GeV) and the steeper spetrum ($N \sim 8-10$). We find modifications of order $\mathcal{O}(40\%)$ and $\mathcal{O}(25\%)$, respectively. Thus, the $R_{AA}$ after BMS evolution retains a significant sensitivity to the initial medium condition, highlighting the need for a framework that consistently captures the smooth transition around $\theta_c$~\cite{Abreu:2024wka}. 

The present work provides a first demonstration of this approach; achieving higher accuracy in the perturbative sector for phenomenological studies will require systematic extensions of the framework. In particular, solving the BMS equation beyond the $R\ll 1$ approximation would improve the treatment of soft-collinear geometry. The framework developed here 
naturally admits systematic order-by-order pQCD improvements, including higher-order corrections through the BMS equation, currently known up to three loops~\cite{Caron-Huot:2016tzz}, and finite-$N_c$ effects via its duality with the JIMWLK equation~\cite{Hatta:2013iba,Hagiwara:2015bia}.
On the medium side, which is fully factorized and enters only through the initial condition, $S_{1,\rm med}$ and $S_{12,\rm med}$ can be calculated within more realistic descriptions of the medium, relaxing some of the hypothesis used in Eqs.\,\eqref{eq:P1med}-\eqref{eq:main-result}~\cite{Armesto:2011ir,Abreu:2024wka,Kuzmin:2025fyu} and incorporating pQCD corrections~\cite{Arnold:2020uzm} at arbitrarily high orders for a weakly coupled quark-gluon plasma as well, including medium-induced parton cascade effects and medium response \cite{Mehtar-Tani:2024mvl,Mehtar-Tani:2022zwf,Soudi:2026fls,Soudi:2025lei}. This work thus paves the way for a systematic phenomenology of high-$p_T$ jet suppression, from boosted color-singlet probes~\cite{Apolinario:2017sob} to soft-collinear resummation for jet substructure and energy-energy correlators in heavy-ion collisions~\cite{Andres:2022ovj}.

\vspace{0.5cm}

\begin{acknowledgments}
\noindent{\bf Acknowledgements.}
P.~C. is funded by the Agence Nationale de la Recherche under
grant ANR-25-CE31-5230 (TMD-SAT). K. E. was supported in part by the U.S. Department of Energy, Office of Science, Office of Workforce Development for Teachers and Scientists (WDTS) under the Science Undergraduate Laboratory Internships Program (SULI). Y.~M.~T. was supported by the U.S. Department of Energy under Contract No. DE-SC0012704. We are grateful for the support of the Saturated Glue (SURGE) Topical Theory Collaboration, funded by the U.S. Department of Energy, Office of Science, Office of Nuclear Physics. 
\end{acknowledgments}
\bibliographystyle{apsrev4-1}

\bibliography{references.bib}

\section{Supplemental material}

\subsection{Threshold Factorization of the Jet Cross Section}
The observable of interest is the inclusive jet production cross-section which in the small cone size approximation factorizes as 
\beq \label{eq:jet-xsect}
\frac{\rmd  \sigma }{\rmd p_T } =\sum_{i=q,g} \int \rmd z H_i(p_T/z,\mu) J_i(z,\mu/Rp_T)\,.
\eeq
Here, $H_i$ denotes the hard matrix element for the production of a jet of flavor $i$ at the scale $\mu$, while $J_i$ is the jet function that encodes the collinear fragmentation of partons along the jet direction. The central working assumption of this proposal is that the short-distance jet production process factorizes from its subsequent interaction with the extended QGP environment. Under this assumption, nuclear effects enter the hard function primarily through modifications of the nuclear parton distribution functions (nPDFs). More importantly, medium-induced effects are incorporated through the long-distance dynamics encoded in the jet function $J_i$. Unlike in proton--proton collisions, where the jet function is evaluated in the vacuum, the relevant matrix elements in heavy-ion collisions must be evaluated in the quantum state of the medium,
\begin{equation}
J(z)\sim \langle 0|\mathcal{O}(z)|0\rangle
\;\longrightarrow\;
\langle \mathrm{med}|\mathcal{O}(z)|\mathrm{med}\rangle\,,
\end{equation}
where $\mathcal{O}$ is a bilinear operator constructed from collinear fields. This replacement captures the modification of jet evolution arising from interactions with the medium while preserving the factorized description of the hard production process.

A second approximation consists of taking the so-called threshold limit of the jet function, $z\to 1$ \cite{Dai:2017dpc,Liu:2017pbb}. This approximation is motivated by the observation that the inclusive jet spectrum, and consequently the hard function, falls steeply with transverse momentum, $H(p_T)\sim p_T^{-N},$
with $N\gg 1$. In this regime, the dominant contribution arises from configurations in which the observed jet carries nearly all of the momentum of the initiating parton. Up to corrections suppressed by powers of $1/N$, the jet function factorizes further into a hard-collinear matching coefficient, which encodes the vacuum dynamics associated with energetic collinear radiation, and a collinear-soft function built from correlators of Wilson lines. 
\begin{align}\label{eq:cs-factorization}
& J(z,R,\mu) =\int_0^\infty \rmd\epsilon \, \delta((1-z)E-\epsilon)\,\sum_{m=1}^\infty \prod_{i=1}^m \int\frac{\rmd \Omega_i}{4\pi} \nn &C^{(m)}_{\rm coll}(\{n_i\},\mu) S^{(m)}_{\rm soft}(\{n_i\},\epsilon,\mu) +\cO(1/N)\,, 
\end{align}
where $m$ denotes the number of resolved collinear partons inside the jet and $\{n_i\}\equiv\{n_1,\cdots, n_m\}$ denote the $m$ collinear directions. 
The collinear-soft operators are defined in terms of Wilson-line correlators along the directions of the collinear partons,
\begin{align}\label{eq:soft-function-m}
&S^{(m)} (\epsilon,R) \equiv  \sum_{X} \,\Theta_{\rm alg}\, \delta\!\left(\epsilon - \bar n \cdot p_{\rm out} \right) \nn& 
\langle \med| U_m^\dag \cdots U_1^\dag U_0^\dag | X \rangle  
\langle X| U_0  U_{1}\cdots U_m | \med \rangle \,,
\end{align}
where it is convenient to study in the light cone gauge $A^+=0$ such that $U_0=1$ and radiation from the medium color charges moving in the anti-collinear direction are suppressed \cite{Gelis:2005pt,Mehtar-Tani:2006vpj}. 

In this work we focus on the leading logarithm $\ln (1-z) \sim \ln N$ order. With the choice $\mu\sim p_T$ in order to fully absorb the large logarithms in the soft function it suffices to consider only the one-body matrix element in the factorization formula: 
\beq 
J(z,R,\mu) \simeq \int_0^\infty \rmd\omega \, \delta((1-z)E-\omega)\,C^{(1)}_{\rm coll}(\mu) S^{(1)}_{\rm soft}(\epsilon,\mu)\,.\nn
\eeq
which leads us to the central object of this work 
\begin{align}
&S^{(1)}_{\rm soft}(\epsilon,\mu)= \sum_{X} \,\Theta_{\rm alg}\, \delta\!\left(\epsilon - \bar n \cdot p_{\rm out} \right) \nn& \langle \med| U_1^\dag  | X \rangle  
\langle X|  U_{1} | \med \rangle \,.
\end{align}
Recall that the cross-section depends on the jet radius $R$, which defines "in" and "out" regions in phase space for gluon radiation. 

$S^{(1)}_{\rm soft}(\epsilon)$ acquires the interpretation of a probability distribution for a parton of flavour $i$ and transverse momentum $p_{T}+\varepsilon$ to lose an energy $\varepsilon$ into the out region defined by the phase space outside of its jet.
Making further use of the steeply falling hard partonic cross-section, the factorization formula above can conveniently be rewritten in Laplace space. Up to power corrections in $N\varepsilon^2/p_T^2\sim 1/N$, one obtains the factorization \eqn{eq:factorization-LL}.

\subsection{Asymptotics of Color Coherence} 

\begin{figure}
    \centering
    \includegraphics[width=0.95\linewidth,page=2]{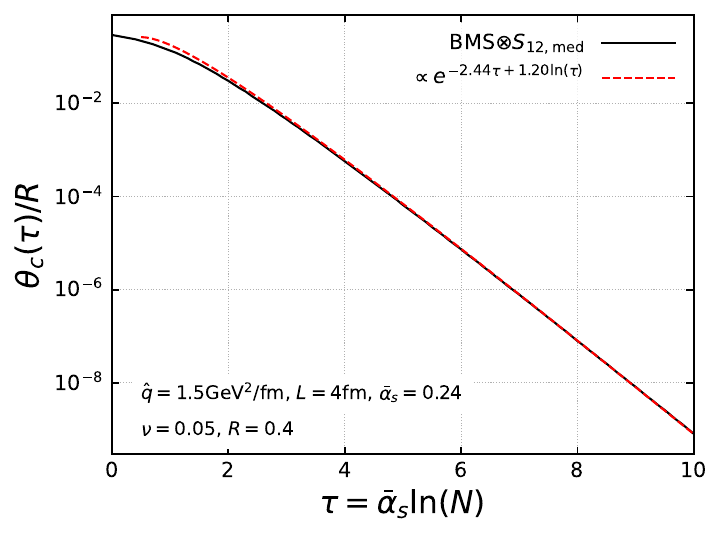}
    \caption{Energy dependence of the coherence angle $\theta_c$.}
    \label{fig:thetac-tau}
\end{figure}

In this Appendix, we provide a numerical crosscheck of the asymptotic scaling behavior of the coherence angle given by Eq.\,\eqref{eq:log-thetac} after BMS evolution.
This $\tau$-dependence of $\theta_c$ in the medium is displayed in Fig.\,\ref{fig:thetac-tau}. The black curve is the result of the numerical solution of Eqs.\,\eqref{eq:bms-1}-\eqref{eq:bms-12} with Eqs.\,\eqref{eq:P1med}-\eqref{eq:main-result} providing the initial condition. $\theta_c(\tau)$ is defined as the angle for which $S_{12}(\tau,\theta_{12})$ is equal to 1/2. On the other hand, the red dashed curve is the exact analytic asymptote Eq.\,\eqref{eq:log-thetac} based on the duality between the BK and BMS equations. The two curves agrees very well for $\tau\gtrsim 5$. One can also notice that for phenomenological values of $\tau\sim 0.2-0.5$, the vacuum dynamics has not yet reached its universal regime meaning that the inclusive jet spectrum remains sensitive to the initial condition. 

\subsection{Numerical setup}

Finally, we provide in this Appendix more details on the numerical resolution of Eqs.\,\eqref{eq:bms-1}-\eqref{eq:bms-12} in the main text. First, we rescale all two dimensional vectors by $R$, such that one can set $R=1$ in Eq.\,\eqref{eq:bms-12}. We then rely on the "hidden symmetry" of the BMS equation in the single cone case discussed in~\cite{Hatta:2009nd} in order to express $S_{12}(\mu,\nu,\boldsymbol{\theta}_1,\boldsymbol{\theta}_2)$ as a function $\tilde S(\mu,\nu,x)$ of the variable $x$ defined as
\begin{align}
    x&=\sqrt{\frac{d_{12}^2}{1+d_{12}^2}}\,,\quad 
    d_{12}^2=\frac{\boldsymbol{\theta}_{12}^2}{(1-\boldsymbol{\theta}_1^2)(1-\boldsymbol{\theta}_2^2)}\,,
\end{align}
such that
\begin{align}
    S_{12}(p_T,\nu,\boldsymbol{\theta}_1,\boldsymbol{\theta}_2)&=\tilde S_{12}(p_T,\nu,x)\,.
\end{align}
In terms of $\tilde S$ and $\tau=\bar\alpha_s\ln(\mu\nu)$, Eq.\,\eqref{eq:bms-1}-\eqref{eq:bms-12} become
\begin{widetext}
\begin{align}
    \frac{\partial \tilde S_{12}(\tau,x)}{\partial\tau}&=-\frac{1}{2}\ln\left(1+\frac{x^2}{1-x^2}\right)\tilde S_{12}(\tau,x)+\int_{|\boldsymbol{\theta}'|<1}\frac{\rmd^2\boldsymbol{\theta}'}{2\pi}\frac{x^2}{\boldsymbol{\theta}'^2(x^2+\boldsymbol{\theta}'^2-2x|\boldsymbol{\theta}'|\cos(\phi'))}\nonumber\\
    &\times \left[\tilde S_{12}\left(\tau,\sqrt{\frac{x^2+\boldsymbol{\theta}'^2-2x|\boldsymbol{\theta}'|\cos(\phi')}{1+x^2|\boldsymbol{\theta}'|^2-2x|\boldsymbol{\theta}'|\cos(\phi')}}\right)\tilde S_{12}(\tau,|\boldsymbol{\theta}'|)-\tilde S_{12}(\tau,x)\right]\,,\\
        \frac{\partial \tilde S_1(\tau,x)}{\partial \tau}&=\left[\ln(R/2)+\frac{1}{2}\ln(1-x^2)\right]\tilde S_1(\tau,x)+\int_{|\boldsymbol{\theta}'|<1} \frac{\rmd^2\boldsymbol{\theta}'}{2\pi}\frac{1}{x^2+\boldsymbol{\theta}'^2-2x|\boldsymbol{\theta}'|\cos(\phi')}\nonumber\\
    &\times \left[\tilde S_{12}\left(\tau,\sqrt{\frac{x^2+\boldsymbol{\theta}'^2-2x|\boldsymbol{\theta}'|\cos(\phi')}{1+x^2|\boldsymbol{\theta}'|^2-2x|\boldsymbol{\theta}'|\cos(\phi')}}\right)\tilde S_1(\tau,|\boldsymbol{\theta}'|)-\tilde S_1(\tau,x)\right]\,.
\end{align}
\end{widetext}
In the equation satisfied by $\tilde S_1$, we have also factored out the linear term given by $\bar\alpha_s\ln(\mu\nu)$ in Eq.\,\eqref{eq:bms-1} thanks to the substitution $\tilde S_1\to e^{\tau^2/(2\bar\alpha_s)+\gamma_E\tau}\tilde S_1$.
This coupled set of equations is then solved on a logarithmic grid of $x$ values, consisting of 1500 points between $\ln(x)=-40$ and $\ln(x)=0$. The evolution in $\tau$ is performed using the Euler method, advancing the solution from one $\tau$ step to the next. For Figs.\,\ref{fig:2prong-BMS}-\ref{fig:thetac-tau}, we use a time step of $\Delta\tau = 2\times 10^{-3}$. The integral over $\boldsymbol{\theta'}$ is evaluated using Simpson's rule with 1500 points in $\ln |\boldsymbol{\theta'}|$ and 20 points in the azimuthal angle $\phi'\in[0,\pi]$.

\end{document}